\documentclass[apj]{emulateapj}
\usepackage{graphicx}
\usepackage{amssymb}
\usepackage{amsmath}
\usepackage{natbib}
\usepackage{rotating}
\usepackage{subfigure}
\usepackage{mathtools}
\usepackage{mathbbol}

\def\beq{\begin{equation}}
\def\eeq{\end{equation}}
\def\beqa{\begin{eqnarray}}
\def\eeqa{\end{eqnarray}}
\def\bseq{\begin{subequations}\begin{eqnarray}}
\def\eseq{\end{eqnarray}\end{subequations}}

\def\uvp{{{$uv$\ plane}}}
\def\kperp{k_{\bot}}
\def\kpar{k_{\|}}
\def\k{{\bf k}}

\def\n{{\bf n}}

\def\N{{\textsf{\textbf{N}}}}

\def\th{{\mathbf{\theta}}}

\def\tI{{\tilde \I}}

\def\B{{\textsf{\textbf{B}}}}
\def\Bv{{\tilde{\B}}}

\def\m{{\bf{m}}}

\def\u{{\bf{u}}}

\def\v{{\bf{v}}}

\def\m{{\bf{m}}}
\def\I{{\textit{\textbf{I}}}}

\def\F{{\textsf{\textbf{F}}}}

\def\m{{\bf{m}}}

\def\({{\big(}}
\def\){{\big)}}

\begin{document}
\title{The Fundamental Multi-Baseline Mode-Mixing Foreground in 21~cm EoR Observations}
\author{Bryna J. Hazelton\altaffilmark{1}, Miguel F. Morales\altaffilmark{1}, Ian S. Sullivan\altaffilmark{1}}

\altaffiltext{1}{University of Washington, Seattle, 98195}

\begin{abstract}

The primary challenge for experiments measuring the neutral hydrogen power spectrum from the Epoch of Reionization (EoR) are mode-mixing effects where foregrounds from very bright astrophysical sources interact with the instrument to contaminate the EoR signal. In this paper we identify a new type of mode-mixing that occurs when measurements from non-identical baselines are combined for increased power spectrum sensitivity. This multi-baseline effect dominates the mode-mixing power in our simulations and can contaminate the EoR window, an area in Fourier space previously identified to be relatively free of foreground power. 

\end{abstract}

\maketitle

\section{Introduction}
\label{intro}

Observing redshifted 21 cm neutral hydrogen emission from the Epoch of Reionization (EoR) is an exciting new tool for observational cosmology. These observations have the potential to reveal the timing and duration of reionization, to constrain models of star and galaxy formation and to determine what kinds of sources dominate reionization (for reviews see \citealt{Morales:2010cb, Furlanetto:2006bq}). Several new arrays are under construction or have recently been built to observe the 21~cm power spectrum, including LOFAR (LOw Frequency ARray\footnote{http://www.lofar.org/}), PAPER (Precision Array for Probing the Epoch of Reionization\footnote{http://astro.berkeley.edu/$\sim$dbacker/eor/}) and the MWA (Murchison Widefield Array\footnote{http://www.mwatelescope.org/}).

The major challenge for EoR observations is the presence of astrophysical foregrounds that are 4-5 orders of magnitude brighter than the EoR signal. Early work recognized that the EoR signal could in principal be separated from the astrophysical foregrounds because the foregrounds have smooth spectra and so should be concentrated in the first few line-of-sight Fourier modes ($\kpar$) while the EoR signal would extend up to much higher $\kpar$ modes \citep{Morales:2004bv, Zaldarriaga:2004gf, Jelic:2008tx, Wang:2006gm, Harker:2009tp}. Unfortunately this picture is somewhat complicated by the interaction of the foregrounds with realistic instruments, which have frequency dependent responses. This interaction is called mode-mixing because the chromatic instrumental response can add spectral structure to the astrophysical foregrounds, throwing foreground power to higher $\kpar$ modes and obscuring the EoR signal. A number of recent papers have begun to shed light on these mode-mixing effects through detailed examinations of how foregrounds propagate through the instrument and analysis.

Precision mode-mixing simulations by \citet{Datta:2010he} first identified a distinctive wedge shape in $\k$ space, with the mode-mixing predominantly below a $\kpar \propto \kperp$ line. Investigations of the response of single baselines to flat-spectrum foregrounds by several groups \citep{Morales:2012ja, Trott:2012fb, Vedantham:2012cm, Parsons:2012ke} showed that the wedge shape is characteristic of smooth spectrum astrophysical sources interacting with the chromatic response of the baselines. This mode-mixing occurs because the baseline integrates over a region of the \uvp\ to form visibilities and the baseline length in wavelengths varies with frequency. This results in a slowly varying value for the visibility as the baseline moves through the $uv$ space and this variation bleeds into the line-of-sight direction because of the physical size of the integration region (see \citealt{Morales:2012ja} for a more complete mathematical and pictorial explanation). The wedge shape comes about because longer baselines change length more quickly than shorter baselines, so the power at larger $\kperp$ is thrown into proportionally higher $\kpar$ modes. These authors also identified a region called the ``EoR window'' at low $\kperp$ and high $\kpar$ that should be relatively free of this kind of contamination. 

While these single baseline effects are very important, the mode-mixed power in the \citet{Datta:2010he} simulations is dominated by a multi-baseline effect that we identify in this paper. Multi-baseline mode-mixing occurs when measurements from non-identical baselines are combined together to increase the power spectrum sensitivity, as they are for nearly all proposed observations. This new effect has a shape in $\k$ space that is similar to the previously identified single baseline mode-mixing, with most of the mode-mixed foreground power thrown into the wedge, but it also contaminates the EoR window. The multi-baseline effect is present in all analyses that combine measurements from non-identical baselines or use images to make power spectra.

In the next section we introduce the multi-baseline mode-mixing mechanism using a simplified simulation with only one foreground source and show that it is fundamental to multi-baseline analyses. Then in section \ref{power} we develop a precision simulation of the power spectrum with realistic foregrounds and we conclude in section \ref{conclusion}.

\section{Multi-baseline Mode-Mixing}
\label{overview}

Visibilities from identical baselines (baselines with the same length and orientation) can be added coherently because they measure exactly the same angular modes on the sky. Coherent summing averages down the noise on the measurement, increasing the instrumental sensitivity to the associated modes. Of course baselines that have very different lengths and orientations measure different angular sky modes and cannot be added coherently, but baselines that are similar but not identical are \textit{partially coherent}, that is they measure some of the same modes but they are each sensitive to modes not detected by the other. In the \uvp, baselines integrate over a small region of the plane given by their power response, so the measured visibilities contain signals from a range of $uv$ locations. For identical baselines the integration region is the same, but for partially coherent baselines the integration regions only partially overlap. Partially coherent visibilities contain separate measurements of the modes inside the overlap region, so it is common to coherently add the visibilities in the overlapping region to decrease the noise in those areas. All EoR sensitivity calculations explicitly combine partially coherent baselines \citep{Morales:2004bv, Morales:2005hu, Bowman:2006dx, McQuinn:2006cr, Lidz:2008hg, Beardsley:2012bk}, as the alternative is a dramatic decrease in the potential sensitivity. 

Unfortunately, combining measurements from non-identical baselines also introduces an additional type of mode-mixing. The reconstructed signal at a particular $uv$ point contains co-added signal from an area of the \uvp\ given by the integration areas of all the co-added baselines. Because the locations of the baselines in the \uvp\ vary with frequency, the $uv$ integration pattern for the reconstructed signal at the $uv$ point also varies with frequency, introducing spectral ripple not present in the foregrounds themselves. 

A visibility measurement is described by equation (12) of \citet{Morales:2009ga}
\beqa
\m\([\v,f]\) = \B\([\v,f],[\u,f]\)\F\([\u,f],[\th,f]\)\I\([\th,f]\)  \hspace{10pt} \nonumber \\
	+\ \n\([\v,f]\) \hspace{10pt}
\label{vis_eqn}
\eeqa
which we have adapted to explicitly note the frequency dimension and to match the notation of \citet{Sullivan:2012ev}. This equation describes the integral that is being done by the instrument in the \uvp, the integral of the baseline power response (\B) multiplied by the Fourier transform of the true sky brightness (\F\I), and is mathematically identical to equation (1-13) in \citet{Clark:1999up} up to the thermal noise term (\n). \B\ is only non-zero over a relatively compact region of the \uvp, so it defines the integration region for the baseline. It is important to note that neither the signal nor \B\ is flat over the integration region. The variation in \B\ over that region means that some areas of the \uvp\ contribute more strongly to the visibility than others and the integration over the signal variation leads to information loss on small $uv$ scales (this is the source of the single-baseline mode-mixing described earlier).

For each $uv$ location of interest, visibilities from all the overlapping baselines need to be combined using a weighted average. The reconstructed power at the $uv$ location ($\u_i$) is given by
 \beqa
\tI\([\u_i,f]\) ={\Bv^T([\u_i,f],[\v,f]\)\m\([\v,f]\) \over \Bv^T([\u_i,f],[\v,f]\)\mathbb{1}\([\v,f]\)}   \hspace{10pt}
\label{rec_i_eqn}
\eeqa
where $\Bv$ is the baseline weighting function, $\mathbb{1}$ is a vector of ones and we have dropped the noise term for clarity.\footnote{The noise comes in as a factor of $\N^{-1}([\v,f], [\v,f])$ between the two factors in both the numerator and denominator, where \N\ is the receiver noise covariance matrix. For an interferometer \N\ is diagonal, and at EoR frequencies it is dominated by the sky noise (included in $\Bv$). Since this leads to a nearly constant diagonal $\N^{-1}$ for EoR interferometers, we suppress the term in Equation \ref{rec_i_eqn}.} Substituting equation \ref{vis_eqn} into equation \ref{rec_i_eqn} yields an important term in the numerator: $\Bv^T\B$, which records how much signal from each $uv$ location is mapped to the reconstructed power at $\u_i$. The movement of baselines in the \uvp\ with frequency generates significant frequency dependence in $\Bv^T\B$ because which baselines contribute to $\u_i$, and the locations of those baselines relative to $\u_i$, is strongly frequency dependent. This frequency dependence in the signal mapping generates frequency dependence in $\tI$ even if the true sky brightness is spectrally smooth. Since the power spectra are generated by Fourier transforming in frequency along each $\u_i$, the spectral ripple introduced by combining partially coherent baselines throws power into high $\kpar$ modes.

\subsection{Multi-Baseline Simulation}
\label{simulation}

\begin{figure}
\begin{center}
\includegraphics[width = \columnwidth]{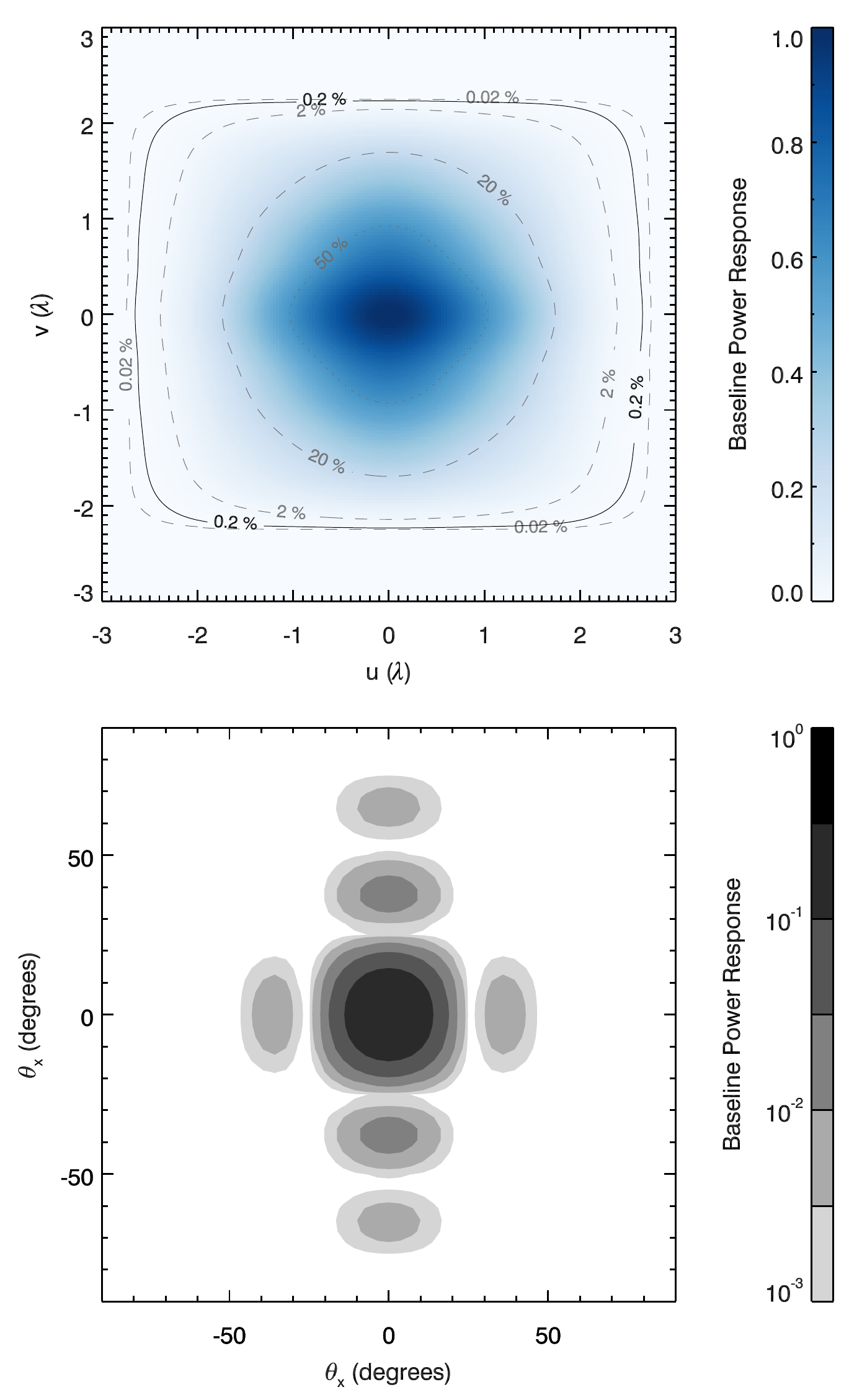}
\caption{The normalized baseline power response ($\B$) for the XX polarization at 150 MHz. \textit{top} The response in the \uvp\ with contours at the 50, 20, 2, 0.2 and 0.02\% levels. The black, solid contour is at the 0.2\% level, the level used in figure \ref{Fig_multibaseline} to indicate the uv integration region. The shape is given by the response function of the square MWA antennas and is asymmetric because it is for the east-west (XX) polarization. \textit{bottom} The corresponding response in the image plane (note the logarithmic color scale).}
\label{Fig_beam}
\end{center}
\end{figure}

To illustrate this effect, we developed precision simulations based on software holography/A projection \citep{Morales:2009ga, Bhatnagar:2008gn} where the visibilities are gridded using the antenna response function for $\Bv$. We use this approach because it can be shown to be information loss-less \citep{Tegmark:1997ct, Tegmark:1997bm}, but the results are general to any analysis that combines measurements from non-identical baselines.  Following \citet{Datta:2010he}, we used the array layout from the originally proposed MWA instrument \citep{Beardsley:2012hq} with 512 square antennas \citep{Tingay:2013ef}. The normalized antenna response ($\B$) for the central frequency used in the simulation is shown in figure \ref{Fig_beam}.

\begin{figure*}
\begin{center}
\includegraphics[height = 8 in]{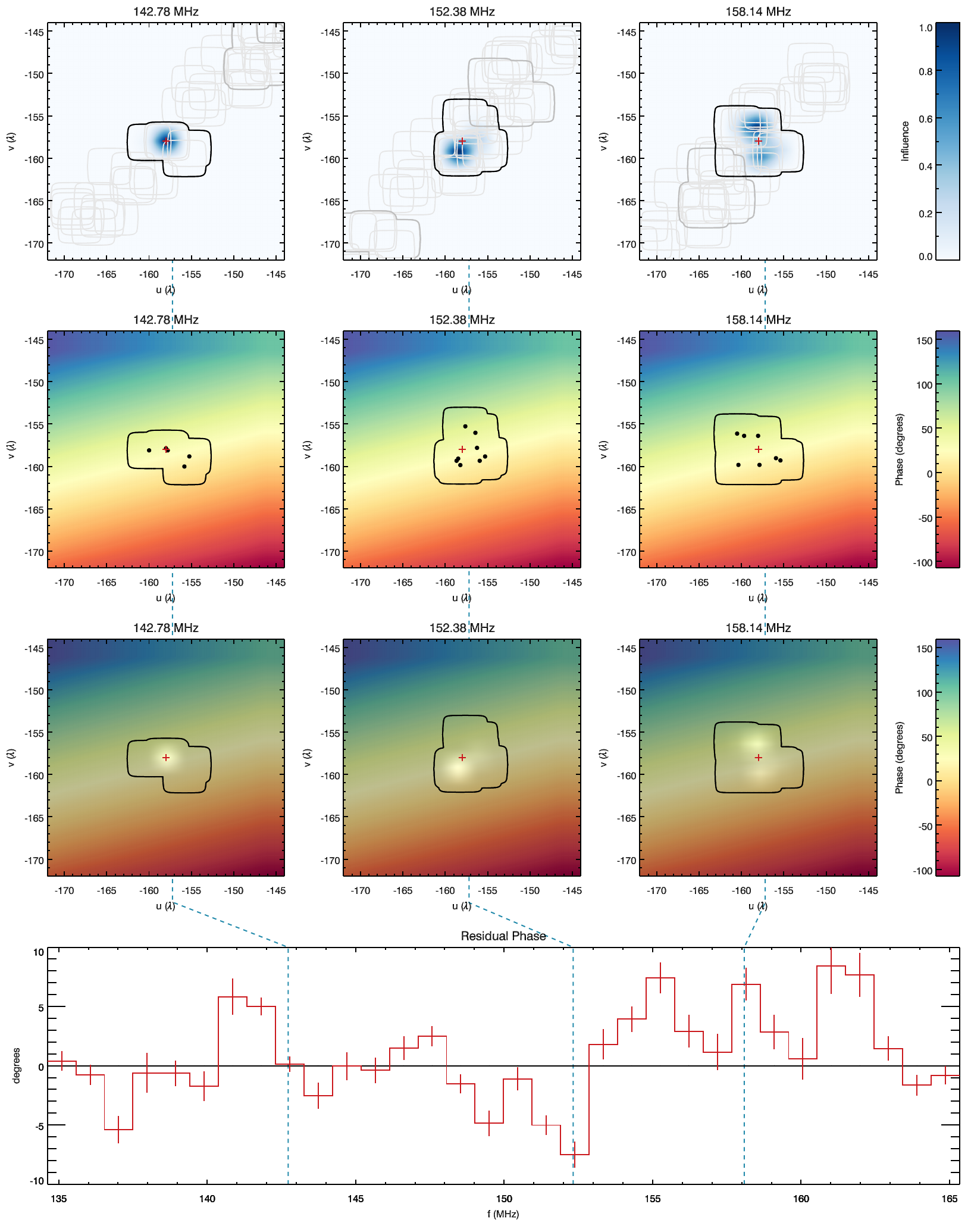}
\caption{This figure demonstrates the effect of multi-baseline mode-mixing in the simplified case of one flat spectrum source at a single location in the \uvp. The \textit{bottom panel} shows the true (flat black line) and reconstructed (red line) phase of the source (relative to the lowest frequency) at the $uv$ location with the selected frequencies marked with blue dashed lines. The upper nine panels zoom into an area around the $uv$ point at the three selected frequencies. The $uv$ point is in the lower left-hand quadrant of the \uvp\ ($uv=0$ is far to the upper right) and is marked by a red cross in each panel. The \textit{top row} shows the integration areas for all the baselines that contribute to the $uv$ point at any frequency in light grey and the set of baselines that contribute at the selected frequencies are outlined in black, defining the region that contributes to the $uv$ point. Baseline lengths increase with frequency, moving to the lower left in subsequent columns, and the integration regions for the other columns are shown in middle grey. The blue heat map shows the relative influence of areas within the integration region on the reconstructed signal. The \textit{second row} shows the true phase of the foreground source with the integration region and the centers of the contributing baselines marked in black. In the \textit{third row} the influence map from the top row is overlaid on the true phase from the second row.  This figure is discussed in detail in section \ref{overview} and there is an animated version showing all the frequency slices in the electronic supplement.
}
\label{Fig_multibaseline}
\end{center}
\end{figure*}

Figure \ref{Fig_multibaseline} demonstrates the effect of multi-baseline mode-mixing using a simplified foreground with a single flat-spectrum point source and focusing on one location in the \uvp. In the bottom panel the true phase of the source at the selected $uv$ location is shown in black (referenced to the lowest frequency). The red line shows the reconstructed phase at that location as a function of frequency, exhibiting obvious frequency structure absent in the true foreground signal. We have selected three representative frequencies, indicated by the dashed blue lines, picking frequencies with almost no phase error, large negative and large positive error (left to right, all frequencies are shown in the animated version in the electronic supplement). Our selected $uv$ point is in the lower left-hand quadrant of the \uvp\ and the upper nine panels zoom into that point, marked by a red cross, at each of the chosen frequencies. 

In the top row, the integration regions  for all the baselines that contribute to the $uv$ point at any frequency are plotted in light grey (using the 0.2\% contour of $\B$). The baselines that contribute at the selected frequency are outlined in black, defining the region of the \uvp\ that contributes to the reconstructed signal at our point. The set of baselines that contribute change with frequency because the baseline length increases with frequency (baselines move to the lower left in subsequent columns) and the integration regions for the other columns are shown in middle grey. The blue heat map shows normalized $\Bv^T\B$ or the relative influence of the areas within the integration region on the reconstructed signal at our $uv$ point. Each visibility is formed by integrating over an area in the \uvp\ with the antenna response function and then the visibilities are weighted by the response at the $uv$ point and averaged to determine the best estimate of the signal there. In other words, the reconstructed signal is a weighted average of the contributing visibilities. The influence map is the effective weight of each location within the integration region on the reconstructed signal.

The second row shows the true phase of our foreground source with the integration region and the centers of the contributing baselines marked in black (each dot corresponds to a grey square in the top row). Finally in the third row we have overlaid the influence map from the top row on the phase from the second row. For the left-hand column the phase is almost perfect because there are two visibilities that are nearly on top of our $uv$ point, dominating the influence map, and the contributions from the other three nearly cancel out. In the center column most of the visibilities are at lower phase and the closest baselines, which dominate the influence map, are clustered just to the left and down from the $uv$ point. In the third column the number of close visibilities above and below the $uv$ point are nearly equal and the influence map is double-lobed but skewed slightly toward higher phase.

The variation in the reconstructed signal with frequency is caused by the dithering of which baselines contribute and how strongly they contribute (i.e.\ the influence map) as a function of frequency. The shape of the multi-baseline mode-mixing ripple is different in each $uv$ pixel and depends on the details of the foregrounds and on the exact locations of the baselines contributing to each location, but unlike the mode-mixing discussed in previous work \citep{Morales:2012ja, Trott:2012fb,Vedantham:2012cm, Parsons:2012ke}, the spatial frequency ($\kpar$) of this ripple is not limited by the field of view of the instrument so it can throw power into higher $\kpar$ modes, including into the EoR window. 

\subsection{Maybe we should....}

While we have used gridding to show the origin of this effect graphically, multi-baseline mode-mixing is inherent in any approach that combines non-identical baselines to create power spectra. However the reasons for this are subtle and it is natural to suggest different analysis approaches to get around the problem. Common ideas include:

\textit{Maybe we should use a different gridding kernel.} The effect of using a different gridding kernel in this case is simply to change the weighting of the different visibilities in the reconstructed signal. However the influence map is a combination of the integral done by the instrument to form the visibility and the gridding kernel, so it won't change dramatically and there will still be shifts in the map that will lead to mode-mixing. In addition, using the antenna response function as the gridding kernel is optimal \citep{Tegmark:1997bm} and using any other gridding kernel would result in a loss of sensitivity to the EoR signal.

\textit{Could we use different antenna elements to decrease this effect?} It is clear that a smoother antenna response function (\B) will cause less power to be thrown to high $\kpar$because the mode-mixed power is windowed with the Fourier transform of the response function. However, all real antennas have a compact integration region in the \uvp\ so the resulting window in  Fourier space is formally infinite. The goal, then, is to increase the volume in k-space over which the EoR signal is greater than the mode-mixed power. Unfortunately there is no clear way to do this. An initial idea might be to increase the size of the antenna elements, making the antenna response smoother and decreasing the extent in $\kpar$ of the mode-mixed wedge. However, if we hold the total collecting area fixed, this will result in a sparser array, which will increase the amplitude of the multi-baseline mode-mixing and decrease the size of the EoR window. Determining the optimal antenna and layout choices will require extensive simulations which will need to include the effects of the observing strategy, the details of the foreground subtraction process and lessons learned from the vagaries of real data.

\textit{What if we changed the visibility weighting to make the influence map frequency independent?} It is not possible to weight the visibilities to achieve a frequency independent influence map because most of the integration is done by the instrument to form the visibilities. For instance in the three maps shown in figure \ref{Fig_multibaseline}, the black outlines show the areas of the \uvp\ that were integrated by instrument. With only 5-8 baselines that overlap the $uv$ point at each frequency, there is clearly no way to weight them to create identical influence maps.

\textit{What about calculating the signal at each $uv$ point without gridding?} This calculation would involve a weighted sum of the visibilities and suffers from the same mode-mixing as in the gridding approach because the locations of the contributing baselines vary with frequency. Indeed the gridding approach is mathematically identical to a weighted-sum of visibilities at each $uv$ point.

\textit{What if we only calculated the power spectrum at the baseline centers?} To measure the power spectrum we need a measurement at each frequency for every $uv$ point we are using. Since the baselines move in $uv$ with frequency, there are no $uv$ locations that have a baseline precisely centered on them at every frequency, leaving no visibilities to use in the analysis. To increase the data we might consider including baselines that are nearly centered on the $uv$ location, effectively setting a radius around the $uv$ point. This scenario puts us right back in the mode-mixing regime, however, because there will be some variation in the locations of the baselines we choose with frequency. The down selection in baselines used in this case would also result in a huge decrease in sensitivity.

\textit{Can we design an array (or chose baselines) to avoid this kind of mode-mixing by having no partially coherent baselines?} Even if there were no baselines that were partially coherent at any frequency, the baselines that contributed to the same location at different frequencies would have different offsets from the $uv$ point, so the influence map would still vary with frequency. In fact our simulations show that the amount of contamination from mode-mixing is decreased in regions with densely spaced baselines and is much worse in areas with very few partially coherent baselines.

Combining a sparse redundant array and the delay spectrum, as pioneered by PAPER \citep{Parsons:2012ke}, does avoid the multi-baseline mode-mixing effect because visibilities from non-identical baselines are never combined and the Fourier transform is along the baseline track through the $uvf$ cube, not along the frequency axis. However, this approach also leads to a significantly smaller EoR window and, for the same collecting area, correspondingly lower EoR sensitivity.

The key issue for multi-baseline analyses using the full EoR window is that the instrument does a weighted integral in the \uvp\ that cannot be reversed and the only way to estimate the signal at any particular $uv$ point is to use the measurements from nearby baselines. Combining these measurements with a weighted average based on the antenna response function (either directly or by gridding) will provide the highest EoR sensitivity, but any approach will suffer from mode-mixing due to the dithering of the contributing baselines.

\section{Power Spectra}
\label{power}

\begin{figure}
\begin{center}
\includegraphics[width = \columnwidth]{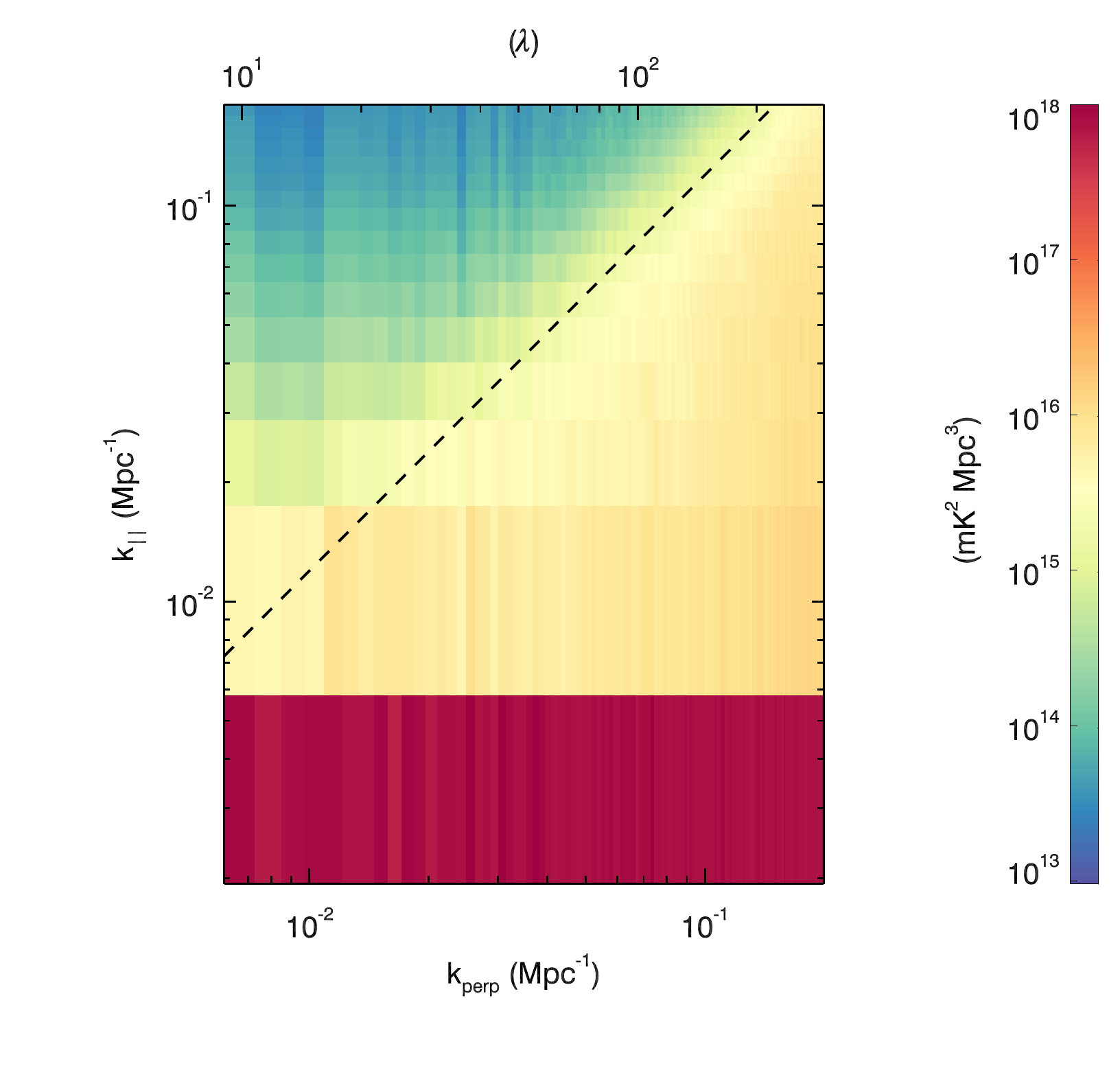}
\caption{Power spectrum showing the mode-mixing contamination in a simulation with approximately five thousand flat spectrum point sources between 0.1 -- 1 Jy (see section \ref{power} for details). The EoR window is above and to the left of the black dashed line.}
\label{Fig_ps}
\end{center}
\end{figure}

In figure \ref{Fig_ps} we show the extent of the mode-mixing contamination in $\k$ space with more realistic foregrounds. The foreground sources for this simulation are flat spectrum with fluxes chosen using the differential source counts from the 6th Cambridge survey at 151 MHz \citep{Hales:1988up}. The sources are restricted to a flux density range of of 0.1 -- 1 Jy, resulting in approximately five thousand sources randomly located in a 30 x 30 deg$^2$ field. The flux range was chosen to represent the dominant sources that would remain after traditional image-based deconvolution of the brightest sources.

This power spectrum was constructed by Fourier transforming in the frequency direction and then calculating a three dimensional power spectrum estimator using weights derived from $\Bv^T\B$. A full description of this estimator will be described in an upcoming paper (Hazelton et al.\ in preparation), but briefly it accounts for the difference in sensitivity of the two Fourier components (i.e.\ sine and cosine) for each mode and marginalizes over the covariance between different modes. The power spectrum is reduced to the two dimensions shown here by doing a variance weighted average along annuli in the $k_x$ and $k_y$ directions.

This power spectrum is similar to those shown in \citet{Datta:2010he}, with a clear wedge shape and a less contaminated EoR window (above the dashed line). The lowest $\kpar$ mode shown is the flat-spectrum ($\kpar = 0$) mode, which would contain all the foreground power if there were no mode-mixing. Our simulations naturally include the single baseline effects identified by \citet{Morales:2012ja}, \citet{Trott:2012fb}, \citet{Vedantham:2012cm} and \citet{Parsons:2012ke} because we form visibilities as described in equation (\ref{vis_eqn}), but we find that the multi-baseline effects dominate the mode-mixed power throughout the \uvp. The contamination that is present in the EoR window is a signature of multi-baseline mode-mixing which can produce contamination to arbitrarily high $\kpar$.

Fortunately there are two factors that mitigate the multi-baseline contamination at low $\kperp$, leaving the mode-mixed power in the EoR window lower than in the wedge. The first effect is that the shorter baselines that contribute at low $\kperp$ move more slowly in $uv$ space with frequency so there is less severe mode-mixing because the distribution of baselines contributing to a particular $uv$ location varies more slowly. In addition, simulations show that areas with denser and smoother $uv$ coverage are less badly affected by this type of mode-mixing, so a dense central core of baselines can significantly reduce the amount of contamination at low $\kperp$.  A dense central core is also desirable for increasing the sensitivity of arrays to the EoR signal, so it is a common feature among most of the arrays currently being built to detect the EoR.

\section{Conclusion}
\label{conclusion}

We have identified a new type of mode-mixing that occurs when measurements from non-identical baselines are combined coherently to increase instrumental sensitivity to the EoR power spectrum. Multi-baseline mode-mixing is present in all analyses that combine measurements from non-identical baselines or use images to make power spectra, and although the majority of the  contamination is in the wedge it also contaminates the EoR window. 

The amplitude of the contamination is significantly lower for regions with smooth, dense $uv$ coverage. It is the dominant mode-mixing term in our simulations using the originally proposed MWA array with 512 antennas, a layout with remarkably smooth $uv$ coverage in its core \citep{Beardsley:2012hq}. While the amplitude of the multi-baseline mode-mixing effect depends on the specifics of each instrument and can only be determined through precision simulations, the dominance for the 512 antenna MWA suggests that it will likely be the primary mode-mixing effect for the first generation of EoR instruments that use multi-baseline analyses. More extensive simulations looking at the dependence of multi-baseline mode-mixing on antenna size, array layout and observing strategy is an important direction for future work and will have implications for the design of future instruments.

Now that we understand the sources of the mode-mixing foreground identified by \citet{Datta:2010he}, the natural question becomes how can we mitigate them. In the next installment of our informal series on the mode-mixing foreground we will explore how to mitigate the multi-baseline mode-mixing foreground. Conceptually, the frequency structure of mode-mixing contamination has a characteristic shape at each $\u_i$ point given by the arrangement of the baselines,  the antenna response,  and the local $uv$ foreground. Unfortunately we cannot just fit for the mode-mixing because the frequency shape depends non-linearly on the foregrounds, but we can solve the problem iteratively given a  good model of the instrument. Because we will only subtract shapes typical of the instrumental response, we will preferentially remove the mode-mixing terms while leaving as much of the EoR signal as possible. In spirit this approach approximates the covariance inversion of \citet{Liu:2011jr} and \citet{Dillon:2013te}, while being computationally straightforward. 

It is not yet clear whether experiments based on single-baseline power spectrum analyses (e.g.\ PAPER) with lower levels of mode-mixing and a smaller EoR window or those opting for the higher sensitivity multi-baseline approach with increased mode-mixing (MWA, LOFAR) will be better able to detect the EoR power spectrum. The outcomes of the currently planned experiments will inform the next generation of Hydrogen Epoch of Reionization Arrays (HERA).

\section*{Acknowledgements}
This work has been supported by the National Science Foundation Astronomy Division through CAREER award 0847753 and NSF Postdoctoral Fellowship 1003314, and by the University of Washington. We wish to particularly thank Danny Jacobs, Adam Beardsley and Aaron Parsons for helpful discussions and the anonymous reviewer whose suggestions greatly improved this paper.

\bibliographystyle{apj}
\bibliography{manual.bib,eor.bib}

\end{document}